%% file: Driver_FracROM.tex
\documentclass[11pt,reqno]{amsproc}
\usepackage[colorinlistoftodos,shadow]{todonotes}
\usepackage{amsmath,amscd,amssymb}
\usepackage[T1]{fontenc}
\usepackage{cite}
\usepackage{color}
\usepackage{graphicx}
\usepackage{psfrag,epsfig}
\usepackage{epstopdf}
\usepackage{multirow}
\usepackage{multicol}
\usepackage{rotating}
\usepackage{fullpage}
\usepackage[letterpaper,margin=1.0in]{geometry}
\usepackage{subfigure}
\usepackage{enumerate}
\usepackage{comment}
\usepackage{lineno}
\usepackage{algorithmic}
\usepackage{algorithm}
\usepackage[small]{caption}
\usepackage[debug=false, colorlinks=true, pdfstartview=FitV, linkcolor=blue, citecolor=blue, urlcolor=blue]{hyperref}
\numberwithin{equation}{section}
\newcommand*\patchAmsMathEnvironmentForLineno[1]{%
  \expandafter\let\csname old#1\expandafter\endcsname\csname #1\endcsname
  \expandafter\let\csname oldend#1\expandafter\endcsname\csname end#1\endcsname
  \renewenvironment{#1}%
     {\linenomath\csname old#1\endcsname}%
     {\csname oldend#1\endcsname\endlinenomath}}%
\newcommand*\patchBothAmsMathEnvironmentsForLineno[1]{%
  \patchAmsMathEnvironmentForLineno{#1}%
  \patchAmsMathEnvironmentForLineno{#1*}}%
\AtBeginDocument{%
\patchBothAmsMathEnvironmentsForLineno{equation}%
\patchBothAmsMathEnvironmentsForLineno{align}%
\patchBothAmsMathEnvironmentsForLineno{flalign}%
\patchBothAmsMathEnvironmentsForLineno{alignat}%
\patchBothAmsMathEnvironmentsForLineno{gather}%
\patchBothAmsMathEnvironmentsForLineno{multline}%
}

\title{Reduced-Order Modeling through Machine Learning Approaches for Brittle Fracture Applications}
\author{A.~Hunter$^{1,*}$, B.~A.~Moore$^{2}$, M.~K.~Mudunuru$^{3}$, V.~T.~Chau$^{3}$, R.~L.~Miller$^{4}$, R.~B.~Tchoua$^{5}$, C.~Nyshadham$^{6}$, S.~Karra$^{3}$, D.~O'Malley$^{3}$, E.~Rougier$^{3}$, H.~S.~Viswanathan$^{3}$, and G.~Srinivasan$^{1}$ \\
{\scriptsize $^{1}$X-Computational Physics Division, Los Alamos National Laboratory, Los Alamos, NM 87545.} \\
{\scriptsize $^{2}$Center for Nonlinear Studies, Los Alamos National Laboratory, Los Alamos, NM 87545.} \\
{\scriptsize $^{3}$Earth and Environmental Sciences Division, Los Alamos National Laboratory, Los Alamos, NM 87545.} \\
{\scriptsize $^{4}$Computer, Computational, and Statistical Sciences Division, Los Alamos National Laboratory, Los Alamos, NM 87545.} \\
{\scriptsize $^{5}$Department of Computer Science, University of Chicago, Chicago, IL 60637.} \\
{\scriptsize $^{6}$Department of Physics and Astronomy, Brigham Young University, Provo, UT 84602.} \\
}
\thanks{$^*$Corresponding author, \texttt{ahunter@lanl.gov}}
\date{\today}
\begin{document}
\maketitle
%
%
\section*{ABSTRACT}
In this paper, five different approaches for reduced-order modeling of brittle fracture in geomaterials, specifically concrete, are presented and compared. 
Four of the five methods rely on machine learning (ML) algorithms to approximate important aspects of the brittle fracture problem. 
In addition to the ML algorithms, each method incorporates different physics-based assumptions in order to reduce the computational complexity while maintaining the physics as much as possible. 
This work specifically focuses on using the ML approaches to model a 2D concrete sample under low strain rate pure tensile loading conditions with 20 preexisting cracks present. 
A high-fidelity finite element-discrete element model is used to both produce a training dataset of 150 simulations and an additional 35 simulations for validation. 
Results from the ML approaches are directly compared against the results from the high-fidelity model. 
Strengths and weaknesses of each approach are discussed and the most important conclusion is that a combination of physics-informed and data-driven features are necessary for emulating the physics of crack propagation, interaction and coalescence.
All of the models presented here have runtimes that are orders of magnitude faster than the original high-fidelity model and pave the path for developing accurate reduced order models that could be used to inform larger length-scale models with important sub-scale physics that often cannot be accounted for due to computational cost.
\newline
\newline
\textbf{Keywords:}~Brittle fracture, machine learning, crack interaction, artificial neural networks.
%
\input{Sections/S1_Intro}
\input{Sections/S2_HOSS}
\input{Sections/S3_ML}
\input{Sections/S4_Results}
\input{Sections/S5_Discussion}
\input{Sections/S6_Summary}

\section*{ACKNOWLEDGMENTS}
The authors would like to acknowledge support from the Los Alamos National Laboratory Directed Research and Development (LDRD) Program through the Directed Research Project \# 20170103DR. 
MKM and SK authors also thank the support of the LANL LDRD Early Career Award 20150693ECR. 
MKM gratefully acknowledges the support of LANL Chick-Keller Postdoctoral Fellowship through Center for Space and Earth Sciences (CSES).
Authors thank the LANL Institutional Computing program for their support in generating data used in this work.

\bibliographystyle{unsrt}
\bibliography{Master_References/paper_bib}
%

\end{document}

%% file: Sections/S1_Intro.tex

\section{INTRODUCTION}
\label{sec:introduction}
Accurately capturing growth, coalescence, and interactions of micro-cracks is critical for determining the failure characteristics of brittle materials used in a variety of applications \cite{Meyer:2000,escobedo2014effect}. 
Fully discretizing each micro-crack within a simulation quickly becomes computationally intractable for component sizes (cm$^3$ and greater) of interest to these applications since each domain can have millions to billions of growing, interacting, and coalescing cracks.  
To overcome such challenges on the macro-continuum scale, modeling approaches either use available experimental data to inform the value of one or more parameter(s) to characterize the fracture mechanism of interest (i.e., an empirical or phenomeonolgical approach), or attempt to describe the failure mechanism within the constitutive description often by focusing on the behavior of a single defect, such as a crack or void, with additional homogenization or statistical averaging to describe the entire evolving ensemble of defects \cite{Shojaei:2013, Zuo:2006, Bazant:1990, Zubelewicz:1987, Needleman:1998, Needleman:2000, Keita:2014}.  
Regardless of the method applied, much detailed information about the evolving damage state in the material is lost, particularly when considering the interaction between cracks.  


In addition, both of these methods for modeling damage evolution require vast information to inform the model development and use.  
This either comes in the form of experimental data, which is often limited due to both cost and loading regimes that can be reached, or knowledge of guiding principles that drives damage evolution, which motivates much of the fracture research occurring at finer length scales.  
Particularly in the latter case, gaining some approximation of how crack evolution drives the overall material response can require tens to hundreds of simulations at lower length scales, which becomes a computational burden for industrial applications.  
Furthermore, this burden increases in order to define uncertainty bounds, which are critical for defining safety factors for commonly used components.  
Hence, reduced-order models that emulate the higher-fidelity models and run in seconds to minutes have great potential to reduce the computational burden and better inform macro-continuum scale models \cite{lucia2004reduced}. 
One way to develop such a reduced-order model is to utilize machine learning (ML) methods and train on the data generated from a high-fidelity model \cite{witten2016data}.

ML is a field that deals with the design, development and implementation of techniques that permit computers to learn based on data \cite{alpaydin2014introduction}. 
The data can come from a variety of sources including measured or inferred data from physical experiments, simulation data from computer models, or a combination of these. 
There have been successful applications of ML across many fields including natural language processing, object recognition, and bio-informatics \cite{Weikum2002,Larranaga2006,Krizhevsky2012}.

ML falls broadly into three categories based on the objective of the technique; unsupervised ML, supervised ML and semi-supervised learning.  
In unsupervised learning, the training data is not labeled, yet the goal is to find underlying structures in the data (such as clusters). Supervised learning is useful when the training data is fully labeled.  
The goal is to find a mathematical function that maps inputs to outputs. 
Examples of supervised ML include random forests (RF), support vector machines (SVM) and artificial neural networks (ANNs) \cite{ho1995random,cortes1995support,mcculloch1943logical}. 
Semi-supervised learning can be thought of as a hybrid method, where one uses a small amount of labeled data with unlabeled data in classification or dimensionality reduction applications. 
This method is used when the cost of placing labels on the data is prohibitive. 

The suitability of an ML technique to a specific application depends both on the goals as well as available data. 
Methods such as decision trees (DT), RF and SVM are generally more effective on small to medium sized datasets, with a few thousand samples or smaller. 
In particular, more complex methods such as ANNs tend to overfit when learning from tens to hundreds of training data points. 
The reverse problem is that of underfitting, when the combination of algorithms, such as DT, and a handful of data points tend to point to very simple models that cannot be generalized successfully in a predictive setting. 

In this research, we have chosen to inform several different ML algorithms using crack network evolution in brittle materials simulations from a validated, finite-discrete element simulator called HOSS \cite{HOSS}.  
When ML is used to replace a high-fidelity model in a predictive setting, supervised learning is an ideal candidate due to the ease with which training data can be generated (and labeled) to span the input space effectively. 
Here we present, compare and contrast five different methods for reproducing aspects of the high fidelity model. 
We narrow our comparison and discussion with two key metrics: prediction of the failure pathway and prediction of the time at which failure occurs.  
These are chosen with the goals to validate the methods, and also highlight which features stand out and must be included in the ML for the method to be successful. 
Since crack interactions are explicitly accounted for in HOSS, a natural way to capture the underlying structure of the crack network is through the use of an equivalent graph representation. 
Some of our models have chosen to exploit this equivalence in order to learn the topological drivers that influence crack propagation using basic concepts such as shortest paths and nearest neighbors. 

This work has an eye toward informing models on larger length scales.  
Ultimately, we aim to inform macro-scale continuum constitutive models with statistical information produced using ML algorithms.  
Such an approach benefits from the computational efficiency that continuum-scale methods have for modeling large specimens and parts (cm and larger), but also has the ability to incorporate sub-scale physics that cannot be resolved in these methods due to computational limitations. 
To the best of our knowledge, this is the first work that looks at building ML-based reduced-order models to describe damage evolution and failure in brittle materials.   

We found that a major challenge with applying ML to predict fracture growth in materials is to engineer the features that can then be generalized to a wide array of problems.  
Among other things, damage evolution in brittle materials is strongly dependent on the applied load, particularly the direction and rate of the load applied.  
To address this problem in fracture mechanics, there are three fundamental fracture modes: Mode I (opening), Mode II (sliding or shearing), and Mode III (tearing) \cite{Sun:2012,Irwin}.  
These three basic fracture modes or their combinations can be used to describe any fracture mode in a cracked body. 
In this study, we have chosen a specific problem with purely tensile quasi-static loading conditions to focus on for training data and model comparison.  
In this case, the primary failure mechanism for failure is Mode I.  
Hence, failure paths are generally perpendicular to the loading direction. 

In Section~\ref{sec:HOSS}, the problem set-up of the high-fidelity simulations and the resulting datasets used for training the ML-based reduced-order models are discussed. 
In Section~\ref{sec:MLapps}, the algorithms for five different ML-based reduced-order models are discussed. 
Validation results of the reduced-order models is presented in Section~\ref{sec:discussion} followed by a discussion on the comparison of the five reduced-order models in Section~\ref{sec:discussion}. 
A summary of the study and key conclusions are presented in Section~\ref{sec:summary}. 

%% file: Sections/S2_HOSS.tex

\section{HIGH-FIDELITY SIMULATIONS}
\label{sec:HOSS}
In this section, we introduce the software suite HOSS, which we have chosen as our high-fidelity model for this work, with specific focus on how HOSS calculates and evolves damage in brittle materials.  
With a clearly defined problem of interest, HOSS was used to generate 150 simulations that constitute the training dataset used by all five ML approaches.  
In addition, HOSS was also used to generate 35 simulations that were used to validate the predictions for failure path and time to failure produced by the ML approaches.  
The details of this dataset and also of the problem set-up are discussed in this section.

\subsection{Hybrid Optimization Software Suite (HOSS)}
For this work, we have chosen to rely on the Hybrid Optimization Software Suite (HOSS) \cite{HOSS1,HOSS3,HOSS2} for our high-fidelity simulations of an evolving crack network, which will be used to both inform and validate the ML approaches described in Section \ref{sec:MLapps}.  
HOSS is a hybrid multi-physics software tool based on the combined finite-discrete element method (FDEM). 
The FDEM approach merges finite element techniques to describe the deformation of the material with discrete element-based transient dynamics, contact detection, and contact interaction solutions, so it can account for both damage evolution, and catastrophic fracture or fragmentation. 
In this section, we briefly review some important aspects of HOSS, as a full description of this model is outside the scope of this article. 
However, we refer the interested reader to the following comprehensive references on the subject for more details: \cite{Munjiza:1992,Munjiza:2004,Munjiza:2012,Munjiza:2015,Rougier:2014,Munjiza:1995}.  
We also note that the work presented here does not depend exclusively on HOSS; other high fidelity modeling approaches that can resolve individual cracks and their evolution could be used for training and validating the ML approaches discussed in later sections.

In the FDEM framework, the solid domains, or discrete elements, are further discretized into finite elements. 
The governing equations are based on conservation of mass, momentum, and energy along with the Newton's laws~\cite{Munjiza:1992,Munjiza:1995}. 
These equation are solved using an explicit central difference time integration scheme~\cite{Rougier:2004}, which makes it necessary to use very small time steps to update the system state dynamically. In HOSS's FDEM framework, cracks form along the boundaries of the finite elements. 
In order to capture fine mechanisms, such as crack nucleation, propagation, branching, reorientation, etc., the crack network must be finely resolved spatially, with dozens to hundreds of finite elements along the length of each crack \cite{Munjiza:2004}. 
As a result, the outputs of simulations involving laboratory sized samples with thousands of incipient micro-cracks can result in petabytes of data.  
In addition, this need for a highly resolved mesh combined with the explicit time integration scheme can result in a need for high performance computing resources for extended durations in order to model the damage evolution and failure of laboratory sized samples and larger.

The problem of interest for this work (described in more detail in the next section) is a 2D sample under a pure tensile load.  HOSS considers two primary modes of failure in 2D: Mode I, which is opening due to tensile load, and Mode II, which is crack growth due to shear loading conditions. 
Since the problem of interest will be dominated by Mode I crack growth, we focus this discussion on the key details as to how HOSS accounts for Mode I crack growth.
However, it must be pointed out that the Mode II crack growth is handled in a similar way to the Mode I case, except different sets of parameters are applied.  
In addition, in a HOSS simulation of our pure tension problem, many of the element edges will not be oriented orthogonally to the applied load.  
Hence, although globally Mode I failure dominates this problem, both shear and opening can occur at a local mesh element scale.

\begin{figure}[!htbp]
  \includegraphics[scale=1.0]{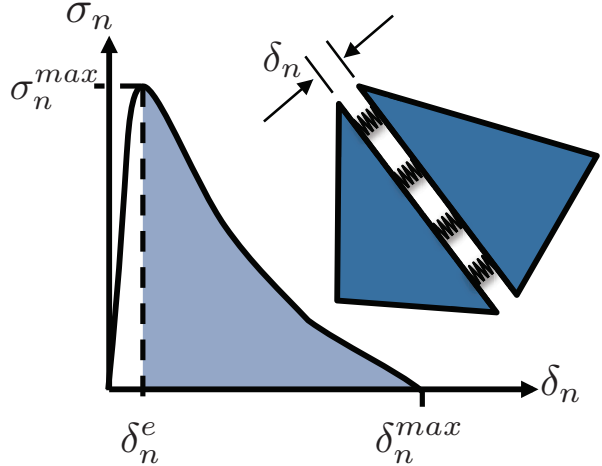}
  \centering
  \caption{Schematic representation of how damage is modeled in HOSS.  
    The cohesive points between element edges are shown for Mode I opening. 
    The amount of displacement between the elements, $\delta_n$, is related to the normal stress, $\sigma_n$, through a similar curve to that shown schematically.  
    The maximum elastic opening of the spring, $\delta_n^e$, and the maximum (elastic and inelastic) opening of the spring, $\delta_n^{max}$, are noted.  
    If $\delta_n>\delta_n^{max}$ the spring is considered to be broken and the stress at which this occurs, $\sigma_n^{max}$ ,is the tensile strength of the material.} 
  \label{fig:one}
\end{figure}

Between the interface of any two finite elements, there lies a user specified number of cohesive points (four is used in all simulations presented and discussed here), which are modeled as springs, as shown in Figure \ref{fig:one}.  
As the two elements undergo tensile load and are pulled apart, the springs within the interface are strained resulting in a small space opening between the elements. 
Similarly, for shear, or Mode II, deformation, there will be four cohesive points that can deform to allow one element to slide relative to another. 
A typical behavior of the springs as a function of the opening created between the edges of two finite elements is shown in Figure \ref{fig:one}.

These cohesive points between elements are responsible for representing the material strength response both in tension and in shear. 
The maximum stress that the spring can withstand in tension is equal to the tensile strength of the material ($\sigma_n^{max}$), while the maximum shear stress that the analogous springs can carry is equal to the shear strength of the material. 
In the first part of the curve ($0<\delta_n\le\delta_n^e$) the springs follow non-linear elastic behavior where no irreversible damage is accrued. 
The springs connecting any two finite elements are very stiff in their non-linear elastic response, i.e., $\delta_n^e \ll\delta_n^{max}$ (the stiffness is two orders of magnitude larger than the Young's modulus of the material) in order to reproduce the response of a continuum medium.  
If the interface between the elements continues to be strained past this elastic limit ($\delta_n^e <\delta_n\le\delta_n^{max}$), the springs enter a strain softening regime which represents the material developing irreversible damage and therefore degrading its strength response. 
Once $\delta_n>\delta_n^{max}$ the spring is considered to be broken and no longer supports any load.  
It is worth noting that the area below the softening portion of the curve, indicated as the shaded region in Figure \ref{fig:one}, represents a strain energy density, i.e., energy per length, dissipated during the fracture process. 
Figure \ref{fig:one} only presents a schematic representation of this curve.  
The actual shape of the curve is found through fitting to experimental results that describe softening in geomaterials \cite{Munjiza:1999}.  

\subsection{Problem Definition, Set-up, and Dataset}
\label{sec:problemsetup}
Since the primary goal of this work is to compare several different ML approaches, we have chosen a single problem to generate the dataset used consistently by all of the ML methods.  
Clearly, we have chosen to focus on material science applications, specifically damage evolution and failure in brittle materials.  
However, for purposes of enabling a direct comparison between ML methods, the key aspect of the dataset is that it is used for training the ML algorithms uniformly across all methods presented in later sections.  

\begin{figure}[!htbp]
  \includegraphics[scale=1.0]{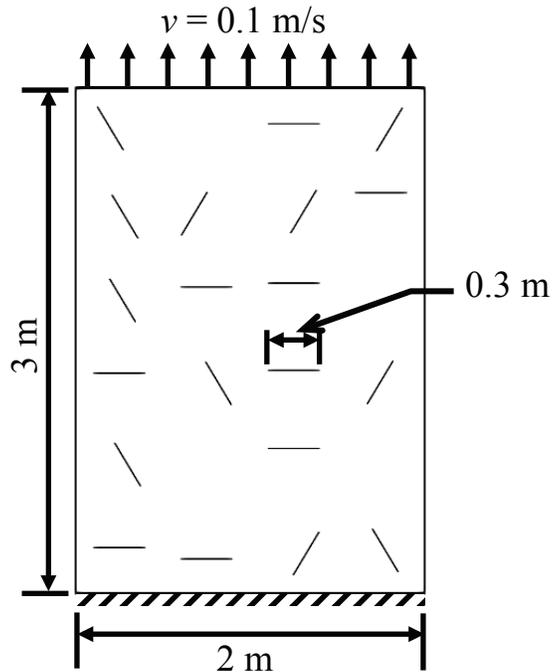}
  \centering
  \caption{Schematic example of a HOSS simulation setup for the problem of interest.  
    Overall 185 HOSS simulations were completed in which the location and orientations (0, 60, or 120 degrees) of the initial cracks were randomly chosen for each simulation.  
    All other geometry, loading, and material parameters were kept the same for all simulations.} 
  \label{fig:two}
\end{figure}

The problem of interest is shown schematically in Figure \ref{fig:two}.  
We have chosen to load a 2D concrete 2m $\times$ 3m specimen in pure tension, hence we expect that Mode I failure will be the dominant material response in all simulations.  
Within the sample, there are 20 initial cracks each with an initial length of 30 cm and three different orientations, 0, 60 and 120 degrees with respect to the bottom of the sample. 
The material is pulled from the top at a constant velocity of 0.1 m/s, and the bottom boundary is fixed. The simulations were run for 700,000 time steps, with a time step of $10^{-8}$s.  
Results from HOSS were output every 2,000 time steps, providing 350 output files of data per simulation.  Each HOSS simulation took about 4 hours of computation time on 400 processors.  
The sample is considered to have failed when a single fracture path connects two opposite boundaries of the material, i.e., a crack spans the entire width of the sample. 
At the point of failure, the material is unable to bear further load, hence we consider this catastrophic failure.  
From a material science standpoint, this simulation configuration imposes similar conditions as a fundamental tension test.  Tension testing has been widely used for material testing, particularly for measuring many material properties including Young's modulus, Poisson's ratio, ductility, yield strength, etc. \cite{nicholas1981tensile}.

The material is assumed to be elastically isotropic in all simulations and the material parameters and boundary conditions used for the concrete sample are: material density ($\rho$): 2500 kg/m$^{3}$; Young's modulus ($E$): $22.6$ GPa; Shear modulus ($G$): $9.1$ GPa; Poisson's ratio ($\nu$): 0.24166; Ultimate Tensile strength ($\sigma_U$): $4.0$ MPa; applied velocity ($v$): $0.1$ m/s at $y=3.0$m and $0$ at all other boundaries.

\begin{figure}[!htbp]
  \includegraphics[scale=1.0]{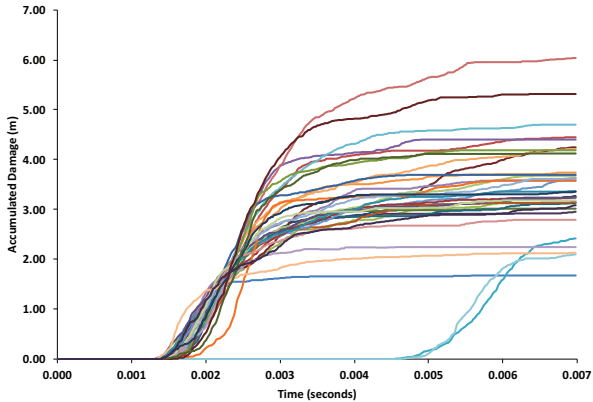}
  \centering
  \caption{Accumulated damage versus simulation time for the 35 simulations in the validation dataset (25 that fail after 0.007 seconds and 10 that have not).  
    Accumulated damage is calculated by determining the amount of distance each crack has propagated every time step and summing to obtain the total distance of new crack.} 
  \label{fig:three}
\end{figure}

Using HOSS, we have run a suite of 185 simulations.  
In each simulation the location and orientations of the initial cracks are randomly chosen.  
Hence, the crack network evolution will be different for every simulation, yet the overall material response should be nominally the same.  
150 of these simulations were used as training data for the ML algorithms discussed in later sections.  
The remaining 35 are used as validation for the ML algorithms.  
Of these 35 simulations, 25 are considered failed after 700,000 time steps and 10 have substantial crack propagation but have not yet failed.  
For the 35 simulations in the validation dataset, the damage accumulation over time as calculated with HOSS is presented in Figure \ref{fig:three}.  
Damage accumulation is calculated by summing the growth of all propagating cracks within the simulation.  
The first thing of note in Figure \ref{fig:three} is that crack propagation does not begin until after some period of loading (about 0.0015s in this case).  
During this time, the material is loading elastically (i.e., $\delta$ is growing from zero to $\delta_n^e$ in Figure \ref{fig:one}).  
Later in the results section (Section \ref{sec:results}), we will see that this time delay until crack growth is difficult to predict when training the ML algorithms unless crack interactions are accounted for. 
This is illustrated through one of the metrics for comparison, time at which failure occurs.

Another metric we will use for comparison of the ML approaches presented in the next section is the location of the failure path.  
Figure \ref{fig:three} shows that the cracks start to propagate very quickly and then level off.  
The majority of simulations level off near 3m, which is the width of the sample.  
As we would expect with a low strain rate, Mode I failure situation, a single large crack will evolve and dominate the growth response.  
Of course other cracks will have some propagation, but in most cases failure is driven by a dominant fracture pathway.  
The ML approaches presented in the next section are trained to predict the location of this dominant fracture path.

%% file: Sections/S3_ML.tex

\section{MACHINE LEARNING APPROACHES}
\label{sec:MLapps}
In this section, we compare five different approaches to modeling the high-fidelity simulations presented in the Section~\ref{sec:HOSS}.  
First, we introduce a shortest path approach that does not rely on ML algorithms and acts a baseline approach for the other methods that do use ML to predict outcomes.  
Each of the following methods incorporate some physically based assumptions \textit{a priori} in order to train the ML approaches to model fracture network evolution.  
The first of these assumes that the component of the crack that is orthogonal to the loading direction is the primary driver of Mode I damage growth, resulting in a method in which the formation of the graph is based on the horizontal (orthogonal to the loading conditions presented above) projections of the cracks.  
This is followed by an approach that assumes that the pairwise interactions between micro-cracks is the primary driver of damage evolution.  
In this case nodes are placed at the center of each crack and edges represent coalescence.   
In the remaining two approaches, the graphs are constructed so that nodes represent crack tips and edges represent the fractured surfaces of the crack.  
In addition, both of these methods are based on the assumption that there is a process zone present at the crack tips, which is a region of localized stress that can result in crack growth.  
The first of these methods uses the idea of of process zone to identify regions of probable catastrophic failure, while the other uses ML to inform the parameters for a simple constitutive model based on the idea of elliptical process zones present at each crack tip.
We will summarize the key details of each method, highlighting the key differences between them.

\subsection{Shortest Path Algorithm (SPA)}
\label{sec:SPA}
This first approach is the simplest application of graph theory principles to exploit the underlying structure of the cracks in the material sample.  
Based on what is known about Mode I opening and our problem of focus, we expect a large horizontal crack to form spanning the width of the sample resulting in failure.  
This method calculates the path across the sample width that requires the least amount of crack propagation.  
In other words, it finds the shortest width-spanning propagation path through a set of preexisting cracks, with the assumption that they will coalesce first because they are closest.  
Crack proximity is taken as the shortest Euclidean distance from a tip of one crack to the body of another.  
Fracture pathways are constructed by joining each pair of nodes with edges weighted according to the Euclidean tip-to-body distance.  
Based on Griffith's criteria \cite{Griffith:1921}, once a crack pair coalesces and becomes the largest crack, it will be the most energetically favorable crack for continued growth.  
In this work, this method is employed without application of ML algorithms and thus is representative of a baseline approach for determining the location of the failure path.  
Because of this, SPA does not take into account information from later times when crack propagation has begun since training data is not utilized.  
Hence, this approach cannot predict the time at which failure occurs for comparison without further enhancement. In this method, each initial crack is represented with a node.  
There are two additional auxiliary ``boundary nodes'' that represent the left and right boundaries of the rectangular sample.  
Graph paths connecting the two auxiliary nodes represent fracture pathways that have spanned the material indicating failure of the specimen.  

The SPA essentially gives an indication of the path of least resistance for crack coalescence forming a large fracture spanning the sample width. 
One of the underlying assumptions for this algorithm is that fracturing only happens along one or two paths out of all possible paths considered. 
This is a reasonable assumption for the quasi-static loading case considered here, since the amount of energy available to break bonds is very low and the system expends the energy in seeking the path of least resistance. 
The initial cracks in the shortest path between boundary nodes is an early-stage estimate of the final failure path (i.e., the path that eventually spans the material sample and causes it to fail, parameterized in terms of the initial cracks it recruited).  

\par\vspace{\baselineskip}
\noindent\fbox{\parbox{0.97\linewidth}{SPA Model Summary:

\begin{enumerate}
  \item Determine the shortest horizontal pathway between the two boundary node connecting as many preexisting cracks as possible.
\end{enumerate}
}}

\subsection{Orthogonal Projection (OP) Approach}
\label{sec:OP}
This method tracks the orthogonal projection of oriented cracks, which for the loading conditions discussed here is also the horizontal projection. 
In essence, we simplify the problem by approximating the orientation dependence of the fracture process with cracks of varying lengths, which is informed by the horizontal projection of the crack.  
This greatly reduces the problem size because we no longer need to keep track of all the individual orientations, which in a realistic material can be infinite.  
However, this approximation assumes that the Mode I failure mechanism has a dominant dependence on the perpendicular component of the cracks, and the additional effect randomly oriented cracks contribute to the evolution of the fracture network can be approximated with ML algorithms.  
If such an approximation produces accurate and reliable results, it would be valuable for macro-scale continuum modeling approaches that seek to average or homogenize the effects of the crack ensemble.  
Since it is too computationally expensive to resolve individual cracks with these methods, assumptions about the dominant orientations or ways to accurately approximate the effects of orientation are valuable for model development.

Unfortunately, the reduction of the orientation dependence also presents some limitations.  
One example of this is some limited ability to predict the failure path, particularly features such as crack branching and crack interaction with vertical neighbors.  
Since all cracks are horizontal and only grow in the direction orthogonal to the applied loading, the predicted failure path will always be a straight horizontal crack spanning the width of the sample, which is of course not a realistic failure path.  
Despite this limitation, the time to failure can still be predicted and the predicted failure path can still provide information about the likely region of failure in the sample.  

This graph is formed by taking the orthogonal (horizontal) projection of the initial micro-cracks provided from HOSS.  
Nodes are assigned to each end of the newly defined horizontal crack network, which now varies in crack length rather than crack orientation.  
Edges are defined in between each node, completing the initial graph.  
Henceforth, the graph representation of the crack network only grows cracks in the horizontal direction, relying on ML algorithms informed with training data from HOSS to account for any differences in the accrued damage due to reducing the orientation dependence.

In the OP approach, the change in crack length $da$ (for a crack of length $2a$) determined through the motion of each crack tip over time is the salient feature evolved and predicted using ML algorithms.  
First, training data is collected from HOSS simulations.  
At every time step, the amount a crack tip moves is calculated and if oriented non-orthogonally to the loading direction a horizontal projection is calculated to determine $da$.  
Polynomial interpolation regression (PIR) was the ML model trained to the data and then used to predict $da$.  
PIR approximates a function with a polynomial of degree $n$ by using ridge regression, more specifically by building a Vandermonde matrix akin to a polynomial kernel~\cite{Ycart:2012}. We used the scikit-learn~\cite{Pedregosa:2011} implementation of this model. 

\par\vspace{\baselineskip}
\noindent\fbox{\parbox{0.97\linewidth}{OP Model Summary:

\begin{enumerate}
  \item Tracks the growth of the horizontal projection of oriented cracks to reduce the problem complexity. 
  \item Learns and predicts the change in crack length $da$ based on $a$ using polynomial interpolation regression (PIR).
 \end{enumerate}
}}

\subsection{Micro-Crack Pair Informed Coalescence (McPIC) Approach}
\label{sec:McPIC}
The McPIC model assumes that pairwise crack interaction and coalescence is the dominant mechanism guiding the overall material response.  
Again due to the computational expense of accounting for discrete cracks and also the analytical complexity of describing multiple cracks, many continuum-scale models do not account for crack interactions at all \cite{Griffith:1921,Paris:1963}.  
Such an assumption will impact the overall material response, particularly the failure time.  
Extracting neighboring pairs of micro-cracks incorporates both the damage accumulation due to crack growth (via coalescence) and the local interaction effects between nearby cracks, connecting these mechanisms to predictions for the global material response.  

\begin{figure}[!htbp]
  \centering
  \includegraphics[scale = 1.0]{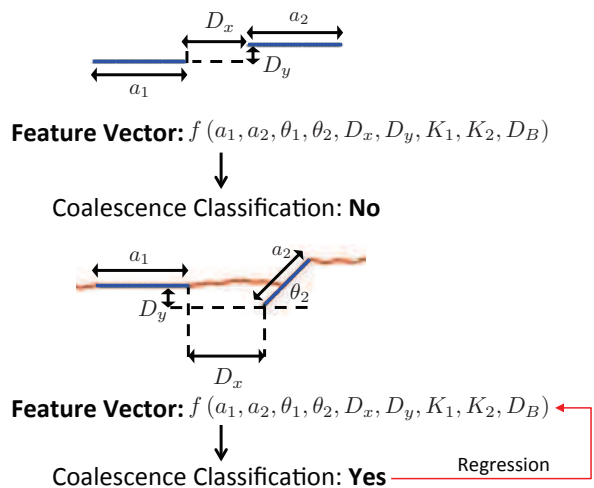}
  \caption{An example of a non-interacting crack pair and a coalesced crack pair.  
    The feature vector includes the crack lengths ($a_1$ and $a_2$), crack orientations ($\theta_1$ and $\theta_2$), the $x$ and $y$ distances between the cracks ($D_x$ and $D_y$), the stress intensity factors at each crack tip ($K_1$ and $K_2$) and the minimum distance to the boundary from either crack ($D_B$). 
    The simulation above has 20 initial fractures which yield 190 data points (with boundary crack-pairs considered).}
  \label{fig:four}
\end{figure}

In the McPIC approach, we isolate all pairs of neighboring micro-cracks and utilize their properties and interactions as training data for the ML algorithms.  
Nodes are placed at the midpoint of existing micro-cracks and edges are formed when two cracks coalesce.  
Each micro-crack pair has a feature vector including geometric information such as, the length of each crack, their respective orientations, and the distance between the two micro-cracks.  
This information must be organized and tracked over time to determine when coalescence occurs and when new crack pairs form.  
A schematic of the coalescence classification process is shown in Figure \ref{fig:four} for both an non-coalescing crack pair and a coalescing crack pair.  
The variables included in the feature vectors are also shown.

Additionally, for the purposes of the ML algorithm, the boundaries of the sample are treated as cracks.  
This is necessary for defining catastrophic failure, which is when crack pairs have coalesced to form a fractured surface spanning the entire sample width.  
This can only occur if a crack pair, in which one crack is a boundary crack, coalesces.  
Hence, the minimum distance from a crack to the nearest boundary, $D_B$, is included in the feature vector.  
These boundary cracks are defined to have zero length, an orientation of 90 degrees and the minimum distance to boundary is set to the distance from the one interior crack to the boundary.  
Treating the left and right boundaries as cracks, each simulation yields 190 unique crack pairs. 
Geometric information about the crack network, and the minimum distance of a crack pair to a boundary can all be extracted from provided initial conditions.  
Whether the cracks within a crack pair coalesce, and when, are the labels for those features.  
The latter information is extracted from the HOSS training simulations and used to train the ML algorithm.

For the results presented in later sections, ANNs are the ML algorithms used in the McPIC approach.  
One concern about using a large network is that if it has more parameters than training datapoints it is prone to overfitting. 
The particular structure of our ANN was relatively small with four layers of 12 neurons, 8 neurons, 4 neurons and 1 neuron in each layer.  
Decreasing the number of neurons from one layer to the next is common practice in ANN creation.  
This structure siphons the number of inputs gradually without large amounts of data loss at any particular stage.

Using ANNs, we are able to make predictions about the materials path to failure (i.e., a sequence of cracks that coalesce amongst each other as well as with the left and right boundaries), and the time that elapses before the material fails.  
Data points were first trained on ANN classifiers which predicted whether crack pairs coalesce. 
All cracks that coalesce were trained on ANN regressions to predict how much time elapses before coalescence occurs. 
Thus, feeding the trained classification and regression networks a set of initial crack pair features would output which cracks coalesce and how long it takes them to coalesce. 
The time to material failure can then be determined by computing the earliest time at which a fracture path that includes both the left and right boundary occurs.  
We note that this method has also been applied using RFs and DTs in place of the ANNs used here.  
Further details on these methods and the comparison across ML algorithms can be found in \cite{Moore:2017}.

Crack pairs are unique and chosen according to their nearest neighbors (Euclidean distance from one crack to another). 
Thus a single micro-crack can be a part of multiple crack pairs.  
This differs from many of the other approaches presented in this section because it does not solely focus on horizontal interactions and propagation.  
Rather, cracks can interact and propagate with neighbors that may be above (vertical propagation) or, more likely, at an angle (combined horizontal and vertical propagation) to any one crack.  
Another potential advantage of this method is that if the pairwise interaction between cracks is modeled accurately, the model could, in principle, be applicable regardless of the sample size, number of initial cracks, or variations in length or orientation distribution, as long as the loading conditions remained constant.  
In addition, it would be expected that the model would extend to other materials within the same material class (i.e., brittle, non-metals).  
This crack pair data implicitly contains a wealth of information related to crack interactions, crack propagation speeds, and the influence of orientations on fracture dynamics, yet is much more concise than the detailed information needed to perform the full HOSS simulation (e.g., high-resolution stress fields, nodal velocities, etc.). 
The computing time and storage of data decreased by 4 and 5 orders of magnitude, respectively, for the previously mentioned 20 crack simulations. 

\par\vspace{\baselineskip}
\noindent\fbox{\parbox{0.97\linewidth}{McPIC Model Summary:

\begin{enumerate}
  \item Focuses on crack pair coalescence and the corresponding time for coalescence to occur. There are two separate but similarly constructed ML models for detecting if there is coalescence and when it occurs.
  \item Feature vectors for each crack pair includes information such as the length of each crack, their respective orientations, the distance between the two micro-cracks, the minimum distance to the boundary from either crack, and the stress intensity factors at the crack tips.
  \item Uses ANNs for the ML algorithm, but can also utilize RF or DT approaches \cite{Moore:2017}.
\end{enumerate}
}}

\subsection{Network-based Fracture Process Zone (NFPZ) Approach}
\label{sec:NFPZ}
In this subsection, we provide a brief description of the network-based fracture process zone (NFPZ)-approach to obtain the most probable failure pathways for dynamic crack propagation under tensile loading. 
This model is based on the idea that a process zone (PZ) exists at each crack tip, where nodes are placed.  
The PZ is a region surrounding the crack tip where a stress concentration resides resulting in damage accumulation as a crack tip propagates over time \cite{Vesely:2007,Hu:1992}. 
Since the stresses in the PZ are very high, very small micro-cracks are formed in the vicinity of the crack tip \cite{Bazant:1987,Bazant2:1990}. 
Over time, as the crack advances, these micro-cracks merge and become a single entity to give continuity to the already existing crack.  
Edges between nodes represent the crack surface, and new edges are formed when the micro-cracks coalesce forming larger cracks. The PZ bridges the cracked and uncracked regions. 
Failure of brittle and quasi-brittle materials under tensile loading typically starts with the development of a PZ around the crack tip \cite{Petersson:1981,Brooks:2013}.
Hence this approach assumes that the most probable path for failure directly correlates with where there are high concentrations of PZs. 

In order to identify regions in which failure is likely to occur, we first identify cracks that are orthogonal to the tensile loading.  
Based on the initial conditions, this corresponds to all of the $0$ degree angle cracks. 
Since these cracks are best oriented for Mode I opening under the loading conditions, we expect that they will grow the fastest. 
Hence, we hypothesize that the likely pathway for the ultimate failure pathway will include one or more $0$ degree angle cracks. 
For each tip of these cracks, we identify the nearest neighboring crack which could interact or coalesce with the horizontal cracks. 
This nearest neighbor can be a $0$ degree, $60$ degree or a $120$ degree crack.   

Interaction and coalescence occurs if the PZs of two neighboring cracks overlap. 
The size of the PZ, $d_{12}$, is given as follows \cite{Wang:1995}
\begin{align}
  \label{Eqn:Crack_Tip_Interaction}
  d_{12} \propto \left(\frac{\sigma}{\sigma_y} \right)^2 (a_1 + a_2)
\end{align}
where $\sigma$ and $\sigma_y$ are the applied and yield stresses of the material, and $a_1$ and $a_2$ are the crack lengths. 
In our case, we assume that the size of the PZ, $d_{12}$, to be 30\% of $(a_1 + a_2)$, that is, $d_{12} = 0.3 (a_1 + a_2)$. 
Hence, if two neighboring cracks fall within this size of PZ, we assume that they are going to coalesce to form a larger crack. 

Once the PZs and potential coalescing cracks have been identified, we next identify the zone(s) of failure. 
Failure zones are regions corresponding to a probably failure pathway, and it is expected that the specimen is going to fail catastrophically in this zone. 
Failure zones contain a set of preexisting crack-pairs that have been identified for coalescence through calculation of the PZs. Through this interaction, there is a strong possibility to form large cracks. 
The large cracks that are formed following coalescence are more favored to propagate compared to smaller cracks \cite{Freund:1990}.  

There may be one or more potential failure zones in a sample, however in a realistic system only one of these pathways will actually correspond to the sample's complete failure. 
Next, we look for weighted shortest paths connecting the sides of the domain that are parallel to the tensile loading direction. 
We impose a constraint that the paths have to traverse through the set of PZ-based large-cracks comprising the failure zones. 
Preexisting cracks are given low edge weights and newly formed cracks are given edge weights based on the Euclidean distance. 
To compute all possible shortest paths for our weighted graph, we use Dijkstra's algorithm \cite{skiena1990dijkstra}.

\par\vspace{\baselineskip}
\noindent\fbox{\parbox{0.97\linewidth}{NFPZ Model Summary:

\begin{enumerate}
  \item Identify the set of preexisting cracks in which PZs at the crack tips overlap and connect them. 
  \item Identify the failure zone. This corresponds to the region which contains larger cracks formed after connecting coalesced preexisting cracks. 
  \item Find the likely failure pathways using Dijkstra path algorithm.
\end{enumerate}
}}

\subsection{Ellipse Process Zone (EPZ) Approach}
\label{sec:EPZ}
The final ML approach presented here is the Ellipse Process Zone (EPZ) model.  
This model utilizes the assumption of a ellipsoidal shaped PZ present at crack tips, similar to the NFPZ model just discussed. 
However, in this approach, a simple constitutive model is developed based on heuristics obtained from observing 2D HOSS simulations.  
We initially assume that the PZ depends on the magnitude of stress intensity at each crack tip. 
The stress intensity factor is a physical quantity used in fracture mechanics to predict the stress state (``stress intensity'') near the tip of a crack caused by a remote load or residual stresses~\cite{anderson2005}. 
To quantify the relationship between stress intensity and material parameters we introduced the crack growth factor ($C^{f})$.  
Our hypothesis is that if we are able to learn this crack growth factor with ML algorithms we can mimic a 2D HOSS simulation. 
We formulate the crack growth factor in terms of the material parameters as follows,

\begin{equation}
  C^{f} \propto \frac{D_y v E \sqrt{a} \cos^{2}(\theta)}{h w \sigma_U \rho D_x} \,.
  \label{eq:Cf}
\end{equation}

The above relation is based on heuristics which relate the $C^{f}$ at a crack tip to different material properties (2D model). 
$C^{f}$ is observed to be directly proportional to Young's modulus ($E$), the square root of the crack length ($a$), the squared cosine of the orientation angle ($\theta$) of the crack taken with respect to the $x$-axis, the resultant velocity due to load applied on the material ($v$), and vertical distance of crack tip from the side the load is applied ($D_y$). 
It is observed to be inversely proportional to the material density, $\rho$, the nearest horizontal distance between crack tip and edge ($D_x$), height ($h$) and width ($w$) of material, ultimate tensile strength ($\sigma_U$).  
Our goal is to improve the crack growth factor using an ML model. In this work we have initially assumed the proportionality factor to be 1. 

In the EPZ model, we represent the initial crack configurations in the 2D material using graph theory principles wherein each crack tip is described as a node and each crack (connected by two nodes) as an edge in graph. 
The graph theory principles in the code are implemented using the \textsf{NetworkX} package \cite{2008_NetworkX_Python}. 
The crack growth factor is calculated at every crack tip or node in the graph at every time step, indicating a propensity for crack growth. 
The dynamic crack propagation is simulated using an algorithm also developed from heuristics. 
The parameters of the algorithm and model are optimized using least squares regression technique as implemented in the \textsf{SciPy} \cite{SciPy} package.  

\begin{figure}[!htbp]
  \centering
  \includegraphics[scale = 1.0]{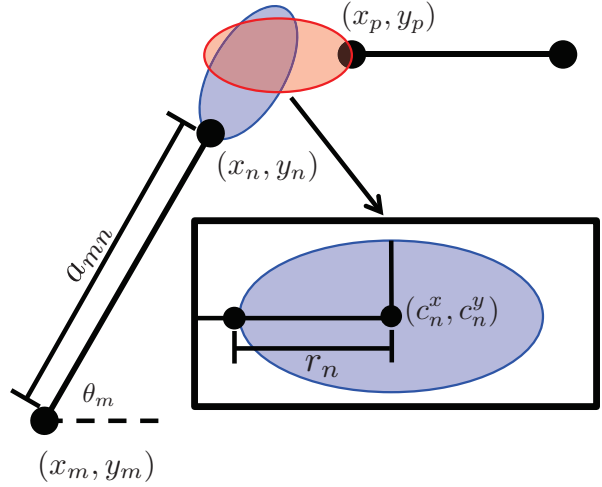}
  \caption{A schematic representation of two cracks and their overlapping PZs as described by the EPZ method.  
    When the elliptical PZs of nodes $n$ or $p$ detect any other nodes (e.g., p or n, respectively), coalescence will occur.}
  \label{fig:five}
\end{figure}

Nodes with a greater propensity for crack growth correlate to crack-tips with larger PZs.  
In addition, crack tips that have a large (small) propensity for growth are included (discarded) from the EPZ calculation.  
By not including nodes that are unlikely to grow, we can further cut the computational expense of this approach.  
For results presented in this work, the cutoff is defined as nodes with a normalized propensity factor, $\bar{P}_{k}$ that is greater than the average of the maximum $C^{f}$ and mean value of $C^{f}$. 
This definition of cutoff was obtained after optimizing the results of EPZ model with many training cases of HOSS simulations.  
Using $C^{f}$, we can calculate the length of the semi-major axis ($r_n$), for the ellipse connected to node $n$ as:

\begin{equation}
  r_{n}= \bar{C}^{f}_{n}*a_{mn}*\gamma\,,
  \label{eq:ma}
\end{equation}

\noindent where the subscripts indicate the node at which the values correspond to, $a_{mn}$ is the crack length spanning nodes $m$ and $n$ (see Figure \ref{fig:five}), and $\gamma$ is a scaling parameter which is updated in a linear fashion from 5.0 to 15.0. 
The linearity constant is learned during training. 
The boundaries 5.0 and 15.0 are also obtained based on observations while training the model.  
Once we determine the length of the semi-major axis, the size and location of the ellipsoidal PZ can be calculated assuming that one vertex is the crack tip (i.e., a node).

Every node is associated with another node through the connection by an edge, as shown in Figure \ref{fig:five}.  
This is representing the crack with two tips at points $(x_n, y_n)$ and $(x_m, y_m)$, which indicates the location of nodes $n$ and $m$ that share an edge.  
Next, to determine the location of the center of the ellipse $(c_n^x, c_n^y)$ with a vertex at a node, say for example $(x_n, y_n)$, we first compute a normalized vector ($\mathbf{N}$) in the direction of orientation $\theta = \theta_n =\theta_{m}$ of the crack. 
The  vector $\mathbf{N}$ is computed as:

\begin{equation}
  \mathbf{N}=\frac{(x_{m},y_{m})-(x_{n},y_{n})}{||(x_{m},y_{m})-(x_{n},y_{n})||_{2}}
  \label{eq:Delta}
\end{equation}

\noindent where  $||(x_{m},y_{m})-(x_{n},y_{n})||_{2} = \sqrt{(x_m-x_n)^2+(y_m-y_n)^2} = a_{mn}$.  
The center of the ellipse located at a distance of $r_{n}$ for node `$n$'  denoted by $(x_n,y_n)$ is then calculated as, 

\begin{equation}
  (c_n^x,c_n^y) = (x_n,y_n) - \mathbf{N} * r_{n}. 
  \label{eq:Delta1}
\end{equation}

\noindent Using the center points and semi-major axis length, we can calculate the ellipse PZ region using the equation for an ellipse given by,

\begin{align}
  &\frac{((x_n-c_{n}^x)\cos(\theta_n)+(y_n-c_{n}^y)\sin(\theta_n))^{2}}{r_n^2} \nonumber \\
  &\hspace{1cm}+\frac{((x_n-c_{n}^x)\sin(\theta_n)+(y_n-c_{n}^y)\cos(\theta_n))^{2}}{(2*r_n(1-e))^2} = 1,
 \label{eq:ellipse}
\end{align}
		
\noindent where $e$ is the eccentricity of the ellipse optimized during training and $\theta$ is the orientation of the crack with respect to the x-axis in the global coordinate system of the 2D sample.  
The orientation, $\theta$ is a dynamic quantity and is computed at each time step. 
In order to determine the orientation $\theta$ for a given crack tip located at $(x_n,y_n)$, we first check at the present time step if there is any other crack tip (node), $(x_p,y_p)$ within the ellipse field (PZ) of influence of $(x_n,y_n)$. 
If so, we assume that the two crack tips attract each other and, hence, move towards each other. 
If so, then we propagate the tip (node) at $(x_n,y_n)$ along the direction of node  $(x_p,y_p)$. 
Then the orientation, $\theta_n$ of the crack tip $(x_n,y_n)$ is given as,

\begin{equation}
  \theta_n = 180^\circ-\frac{180^\circ\arctan\left(\frac{y_{p}-y_n}{x_{p}-x_n}\right)}{\pi}
\end{equation}

If the orientation is greater than 45 degree, we propagate the crack tip with the averaged orientation based on lengths of cracks corresponding to crack tips  $(x_n,y_n)$ and $(x_p,y_p)$. 
This is given as,

\begin{equation}
\theta_n = \frac{a_{mn} * \theta_n + a_{pq} * \theta_p}{2},
\end{equation}

\noindent where $a_{mn}$, $a_{pq}$ are the crack lengths  and $\theta_n$, $\theta_p$ are orientations corresponding to nodes $(x_n,y_n)$ and $(x_p,y_p)$. 
In this way, a crack can reorient in several time steps to coalesce with the nearby crack.  
If there are no overlapping ellipse PZs, then the crack is assumed to propagate orthogonal to the loading conditions (horizontally in the case discussed here). 
The algorithm for orientation assumes Mode I failure, so cracks are expected to propagate orthogonally (horizontally to the vertically applied load in this case) to the applied load.  
However from many training examples, we observed that if the crack's orientation deviates too far from the orthogonal direction, it will curve over several time steps until it can start propagating in the perpendicular direction to the applied load. 
To account for this, we propagate these cracks with half of the current orientation until the crack is oriented horizontally.
 
Once the ellipse PZs are determined, we can evolve the graph.  
The propagation is done by creating a new daughter node at a distance of $dL_n\propto(1+ a_{mn})$ from crack tip $(x_n,y_n)$ in the orientation as described above.  
The constant of proportionality is determined during the training. 
Once the distance and orientation for crack propagation is determined, a new node is created at that location, and the parent node becomes inactive.  
In the case of coalescence, both nodes become inactive and no new nodes are created. In order to detect failure time, at every time step, we check if any nodes have touched the boundaries of sample and if there exists a direct path between two opposite boundaries of the material. 

\par\vspace{\baselineskip}
\noindent\fbox{\parbox{0.97\linewidth}{EPZ Model Summary:

\begin{enumerate}
  \item Developed based on heuristics obtained form observing HOSS simulations.
  \item Simple and based on constitutive relations in physics.
  \item Uses optimization techniques to learn from the data.
\end{enumerate}
}}

%% file: Sections/S4_Results.tex

\section{RESULTS}
\label{sec:results}
For comparison of the five approaches introduced above we have chosen two metrics.  
The first is determination of the failure pathway, followed by prediction of the time at which failure occurs.  
All the previously discussed methods can provide results for the first metric.  
However, for the time to failure, only the OP, McPIC, and EPZ methods can make predictions.  
These metrics were chosen because they can be easily validated with HOSS results to determine the efficacy of each approach.  
In addition, we can easily compare and contrast methods in order to investigate which physical assumptions lead to the best results and which variables are best modeled with the ML algorithms.

Once a ML method is shown to produce valid results it could potentially be used to inform material models, particularly those at macro-scale continuum length scales where it is computationally intractable to resolve individual cracks and their evolution.  
In order to inform continuum-scale constitutive models, statistical information, such as probability density functions, that describes damage accumulation over time would be needed.  
For this purpose, exact locations of the failure pathways and the coalesced cracks becomes less important since the discrete crack network cannot be accounted for in these types of models.  
Rather, the problem set-up described in Section \ref{sec:problemsetup} would, perhaps, be representative of a single cell or element in a large scale finite element or hydrocode model.  
Predicting when failure occurs and how much damage has accumulated will be important to accurately predict the overall material response, which includes accounting for degraded material properties as damage accumulates (i.e., the softening curve) and when and where failure of the specimen or part occurs (i.e., which elements or cells will fail first).  
Providing such information is subject of current and future research, although our second metric (prediction of failure time) is directly relevant to this goal. 

\subsection{Predicting the Dominant Failure Path}
Of the 35 simulations in the validation dataset, 25 have achieved full material failure at the end of the simulation and 10 have not.  
A successful prediction of the failure pathway is defined when the initial cracks within the fracture spanning the sample are the same in both the HOSS calculated failure path and the path predicted any of the ML approaches.  
In results presented here, partial matches are not considered correct or given any credit.  
In other words, if an ML approach predicts an additional initial crack in the failure path or only includes some but not all initial cracks that the HOSS result shows, it is considered an incorrect prediction of the failure path.  
We note that including partial matches would significantly improve the statistics.  
In particular, the McPIC and EPZ methods not only predict growth and coalescence of the proposed failure path, but also propagation of other cracks within the system.  
If the overall damage accumulation was considered, these two methods are able to predict potential matches for propagation and coalescence outside of the crack growth that directly leads to material failure.

\begin{figure*}
  \includegraphics[scale=1.0]{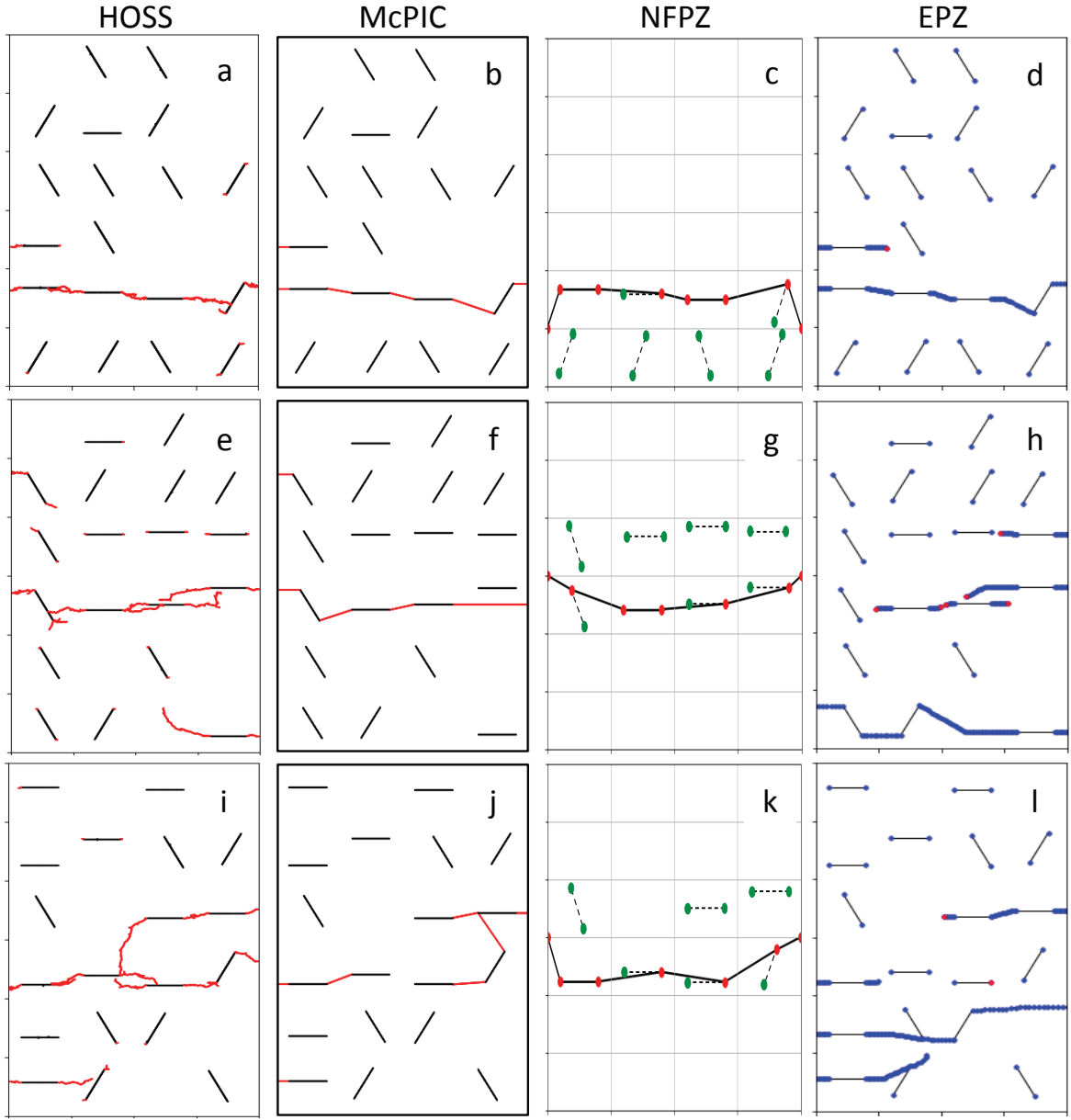}
  \centering
  \caption{Three representative simulation cases showing the HOSS results (Figures (a), (e), and (i)), and the results predicted with the McPIC (Figures (b), (f), and (j)), NFPZ (Figures (c), (g), and (k)), and EPZ (Figures (d), (h), and (l)) approaches.  
    The first row of figures ((a)-(d)) shows an example of a simulation in which the material fully failed.  
    Figures (e)-(h) shows an example of a simulation in which material failure was not achieved.  
    Finally the third row, Figures (i)-(l), are representative of simulations in which the sample failed through crack branching.} 
  \label{fig:six}
\end{figure*}

The dataset has been divided into three categories:~simulations that have achieved full material failure without the dominant fracture pathway branching (20 simulations total), simulations in which the material failed through crack branching of the dominant fracture pathway (5 simulations total), and simulations that did not achieve full material failure (10 simulations total).  
In the latter case, the failure path is defined as the crack path that encompasses the largest percentage of the sample width.  Figure \ref{fig:six} shows a representative example for each of the three categories.  
The first row, Figures \ref{fig:six}(a)-(d), present results for a simulation in which the material failed without any crack branching.  
The second row, Figures \ref{fig:six}(e)-(h), shows an example of a simulation where the material sample is near failure, but the dominant crack pathways did not fully coalesce before the end of the simulation.  
Finally, Figures \ref{fig:six}(i)-(l) in the third row show an example of a sample that failed with a pathway that branched about halfway across the sample.  
The first column (Figures  \ref{fig:six} (a), (e) and (f)) show the HOSS results for each of the three different simulations.  
The remaining three columns present the predicted results from the McPIC, NFPZ, and EPZ methods, respectively. 
Figures from the SPA method are not included since this approach will not deviate from the shortest possible pathway possible based on horizontal distances between preexisting cracks.  
Similarly, figures from the OP method are also not included, since the failure pathway predictions will always consist of horizontal lines spanning the sample width.  

The first row of figures, Figures \ref{fig:six}(a)-(d), is perhaps the most straightforward of the three cases showing a failed sample via a pathway that does not include any crack branching.  
Figures from the McPIC, NFPZ and EPZ methods show that all three of these approaches correctly predict the failure path.  
Although no figures are shown, the SPA and OP approaches are also successful in predicting the failure path for this simulation.  
It is also worth noting that the McPIC and EPZ methods also capture the additional propagation of the crack above the dominant failure path.  

In the second row, Figures \ref{fig:six}(e)-(h) show images corresponding to a sample that has not yet fully failed.  
In this case, the McPIC, NFPZ and EPZ methods all predict that the sample has failed even though the HOSS results show that this is not the case.  
In nearly all of the 10 simulations in which the sample has not yet failed, all the approaches predict early failure.  
This could indicate that there is a key feature missing in these algorithms related to the crack growth rate.  
Of the three images, the most accurate prediction of the likely failure path is produced by the NFPZ approach.  
The McPIC approach also produces promising results for this case.  
The EPZ approach predicts failure along the bottom of the sample, which is incorrect in this case.  
However, this method does show propagation throughout the sample, much of which corresponds to propagation seen in the HOSS results.  
The EPZ approach also produces crack evolution that best reflects crack curvature seen in the HOSS results.  

Finally, Figures \ref{fig:six}(i)-(l) present a sample that failed through crack branching.  
Of the three methods shown in the figure, the NFPZ method best captures the failure path, but does not account for any crack branching behavior.  
The SPA approach predicted a similar pathway to the NFPZ approach, but is also unable to account for branching behavior.  
It is worth nothing the while the McPIC did not correctly predict failure of the sample or the the location of the branching, it does predict a branched configuration.  
Crack branching highlights the importance of accounting for crack-to-crack interactions.  
Since the McPIC method has focused training resources toward all possible crack pairs it is more likely to capture branching mechanisms.  
The incorrect placement of the branch may highlight a need to consider more that just pairwise interactions between the cracks contained within the fracture network.

The overall number of correct predictions made by each method for each of the three categories are shown in Table \ref{table:failurepath}. 
From Table \ref{table:failurepath} we see that the failure path in simulations in which the material fully failed without any crack branching were most often predicted correctly across the methods, with the OP and McPIC methods providing the most correct, exact matches with the HOSS simulations.  
It must be noted that the criterion for a correct match was very strict, expecting better than human performance. 
Full credit was awarded when only the exact cracks predicted by HOSS were replicated by the ML algorithms, and no partial credit was given when the bulk of the path (perhaps except one crack, or including an extra crack) was identified correctly. 
It must be noted that for several of these cases involving branching, the human eye proved quite unreliable in exactly pin-pointing the failure path. 
Given these harsh metrics for success, the fact that our most successful methods predicted the failure path 45\% of the time is highly encouraging.  
The SPA algorithm, which does not utilize an ML algorithm and only used knowledge of initial conditions produced correct exact matches in 40\% of the simulations.  


Since only exact matches in failure path are considered, the number of correct predictions shown in Table \ref{table:failurepath} act somewhat as a lower bound.  
Fracture networks and failure pathways can be extremely complex, and exact matches can be quite difficult to achieve.  
While this was a highly relevant first step towards emulating the expensive HOSS simulations in a fraction of the time, it must be noted that statistics derived from these ML predictions including crack length distributions and overall damage value are aggregated or averaged quantities, and hence more forgiving in their computation. 
These results do provide us with valuable insight as to which methods are best for capturing crack propagation and coalescence, and also indirectly which mechanisms or features are important to include in the ML training information (see Section \ref{sec:discussion}).   

\begin{table*}
  \caption{\label{table:failurepath}The number of correct failure path predictions for the SPA, OP, McPIC, NFPZ, and EPZ methods.  
    A correct prediction of a HOSS failure pathway is defined when the initial cracks within the dominant fracture spanning the sample are the same in the HOSS calculated failure path and the path predicted by the ML approaches. 
    Only exact matches are considered as a correct prediction, although the McPIC and EPZ approaches in particular can also predict other crack propagation.  
    Samples that failed with branching crack pathways are separated into a category since the were the most difficult pathways to accurately predict.  
    When the HOSS simulation is not fully failed, the failure path is defined as the dominant fracture, which is the crack pathway that spans the largest amount of the sample width.}
  \medskip
  \centering
  \begin{tabular}{|c|c|c|c|c|c|c|}	
    \hline
     Description & Number of simulations & SPA & OP & McPIC & NFPZ & EPZ \\
    \hline
     Failed samples without branched pathways & 20 & 8 & 9 & 9 & 6 & 4\\         
     Failed samples with branched pathways & 5 & 0 & 0 & 0 & 0 & 0\\  
     Samples that did not fully fail & 10 & 1 & 2 & 5 & 2 & 1\\  
    \hline
  \end{tabular}
\end{table*}

Table \ref{table:failurepath} also shows that there was not a single exact match for simulations where the material failed via a branched failure path across all the proposed methods.  
Crack curvature and branching is difficult to determine accurately.  
Of the proposed methods, only the McPIC and EPZ would be capable of capturing these failure path configurations.  
A propagating crack in the SPA method, for example, cannot reorient or branch to propagate above or below a dominant crack path.  Similarly, the OP method only considers horizontal propagation of cracks in the material system.  

The final category in Table \ref{table:failurepath} refers to the 10 simulations in the validation dataset that did not achieve full material failure.  
In this case, the McPIC method outperformed the other methods, accurately predicting the dominant fracture pathway 50\% of the time.  
Unlike the cases in which the material is fully failed, this is a significant improvement in comparison to the SPA approach, which highlights the value of including ML algorithms in the approach.  
In the case of nearly failed samples, the ML approaches often underpredict the failure time. 
In other words, the ML approaches predicted a fully formed failure path when the HOSS simulation shows that the sample has not yet completely failed.  
This can be seen in the example simulation comparisons for the McPIC, NFP, and EPZ methods in Figures \ref{fig:six}(e) - (h).

\subsection{Predicting Time to Failure}
\begin{figure*}
  \includegraphics[scale=1.0]{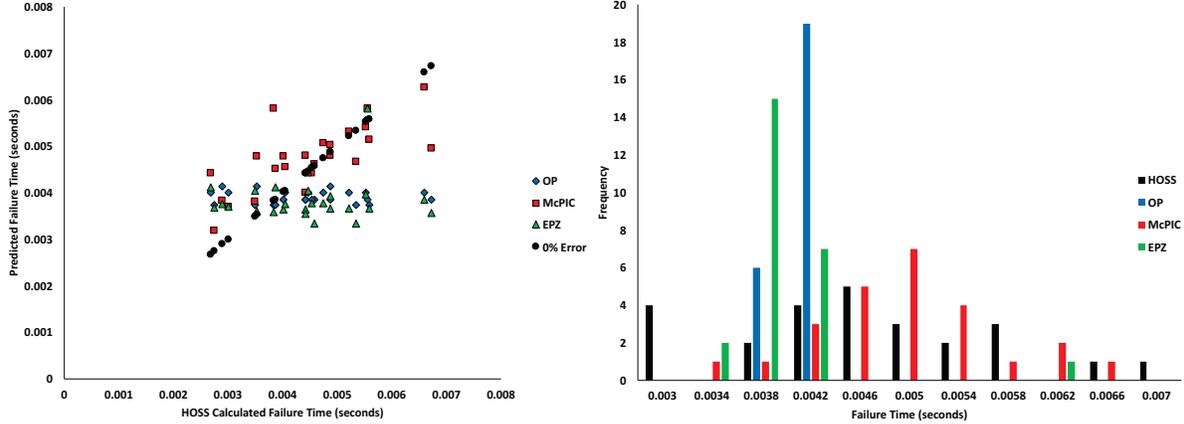}
  \centering
  \caption{Predicted failure times from the OP, McPIC, and EPZ approaches compared to the failure times calculated with HOSS shown   as (a) a scatter parity plot and (b) a histogram.} 
\label{fig:seven}
\end{figure*}

Moving on from predicting the failure path, several methods, namely OP, McPIC and EPZ, can also predict the time at which failure occurs.  
Figure \ref{fig:seven} shows both a scatter plot and histogram of the predicted failure times in comparison to those calculated with HOSS.  
Results are shown for only the 25 simulations that fully failed in the dataset.  
As noted in the previous section, failure time was underpredicted in nearly all cases for the 10 simulations that did not fully fail.

Figure \ref{fig:seven}(a) shows rather narrow bands for the failure time prediction from the OP and EPZ method.  
Essentially these two methods have learned the average time for failure to occur in these types of tensile sample with these loading conditions.  
This is confirmed in Figure \ref{fig:seven}(b), which shows that these two methods predict the failure time for nearly all simulations to be very close to the average failure time calculated by HOSS, 0.0446s.  
It is worth noting that all of the OP and EPZ predictions slightly underpredict the average failure time, with the EPZ method showing a slightly broader spread that the OP method.  
The McPIC method shows a much wider spread, and only fails to capture failure times at the two extremes of the data, which is highlighted in the histogram in Figure \ref{fig:seven}(b). 
Figure \ref{fig:seven}(a) shows that the results produced by the McPIC method follow the parity line.

%% file: Sections/S5_Discussion.tex

\section{DISCUSSION}
\label{sec:discussion}
Of the methods presented, the McPIC model shows the most promising results for both failure path and time to failure predictions.  
This is the only method that considered crack pairs as the unit of interest, highlighting the importance of capturing these interactions.  
The method not only had a high number of correct exact failure path predictions, but also shows the ability to capture coalescence outside of the dominant failure pathway and crack branching behavior.  
This is the only method that did not purely consider geometric factors, but also included crack intensity factors as variables in the feature vectors for the ML algorithms to train with.  
This likely impacted the both the crack propagation and pathway behavior and, subsequently, the time at which failure was predicted.  
We expect that the model would see improvements if tertiary (or higher order) crack interactions were considered rather than just pairwise interactions.  
The initial preexisting cracks are relatively close to one another, hence higher order interactions may have a notable impact on the growth rate of the crack network.  
If this were explicitly accounted for in the algorithm, improvement in the accuracy of predictions would be expected.

The OP method also produced good results, particularly in calculating the failure pathway.  
For this metric, this model was only outperformed by the McPIC method in the case of predicting the dominant crack pathway in simulations where the sample was not yet failed.  
It also only slightly underpredicted the time at which failure occurs (~0.0042s vs HOSS 0.00446s).  
These results essentially show that the accounting for the orientation dependence with varying crack lengths is a reasonable assumption, as long as average times for failure are adequate for the application.  
However, this assumption is only applicable for Mode I failure, and would not be acceptable for shear or combined loading conditions.  
If interactions with neighboring cracks above and below the horizontal growth path were accounted for in the crack growth rate, time to failure would likely be predicted more accurately.

The EPZ method produced interesting results.  
While this method did not outperform others in a significant way, it was able to capture coalescence of cracks outside of the dominant failure pathway.  
It was also able to capture reasonable crack curvature and potentially also crack branching behavior.  
These behaviors are likely a result of the elliptical PZs that exist at each crack tip and evolve in shape and orientation over time and with the location of the crack tip.  
This provides a large amount of variability in the direction in which cracks can grow, which is something that the other methods do not allow for.  
However in the EPZ method, it seems that much of the crack propagation is overpredicted in comparison to the HOSS results.  
In the EPZ images shown in Figure \ref{fig:six}, many of the cracks that are predicted to propagate also are shown to propagate in the HOSS results, however in the EPZ model they seem to propagate too quickly resulting in incorrect coalescence and sometimes incorrect failure path predictions.  
This perhaps reflects a dependence on one or more material factors within the posited crack growth factor (Equation \ref{eq:Cf}) that is too strong.  
In addition, accounting for stress state information at the crack tips could also impact the crack growth rate, improving results.

The NFPZ approach is somewhat a hybrid of the OP and EPZ models, and consequently it performs better than the EPZ method and not as well as the OP method for failure path prediction.  
This model does assume, based on Mode I failure arguments, that horizontal cracks will grow the fastest similar to the OP method.  This idea is used to determine a zone where failure is likely.  
This method also assumes PZs present at crack tips, similar to the EPZ approach.  
The PZs are used to determine which cracks are likely to coalesce, and this also informs the likely location of failure.  
Unlike the EPZ method, the PZs are not recalculated with time and all crack tips have the same PZ size (which is true as long as $a_1$=$a_2$).  
Adding some variability to the PZ calculation (e.g., dependence on orientation, stress state, etc.) could improve results.  
In addition, the likely region of failure may be too narrowly defined.  
Other methods (specifically the OP and McPIC methods) that did not reduce the system to a local failure zone present better results. 

It is worth noting that the SPA approach did produce reasonable results for calculating the failure path based solely on a shortest fracture pathway, without any use of ML algorithms.  
Of course, similar to the OP method, the SPA method is based on assumptions only applicable for Mode I failure scenarios.  
Future modifications to this algorithm could include weighting the shortest path by geometric as well as physical considerations. 
For example, through the consideration of pairwise interactions the probability of two cracks coalescing could be higher when the distance between tips is small. 
This can be incorporated into the weights while computing shortest paths which may improve results. 

\begin{figure*}
\includegraphics[width=\textwidth]{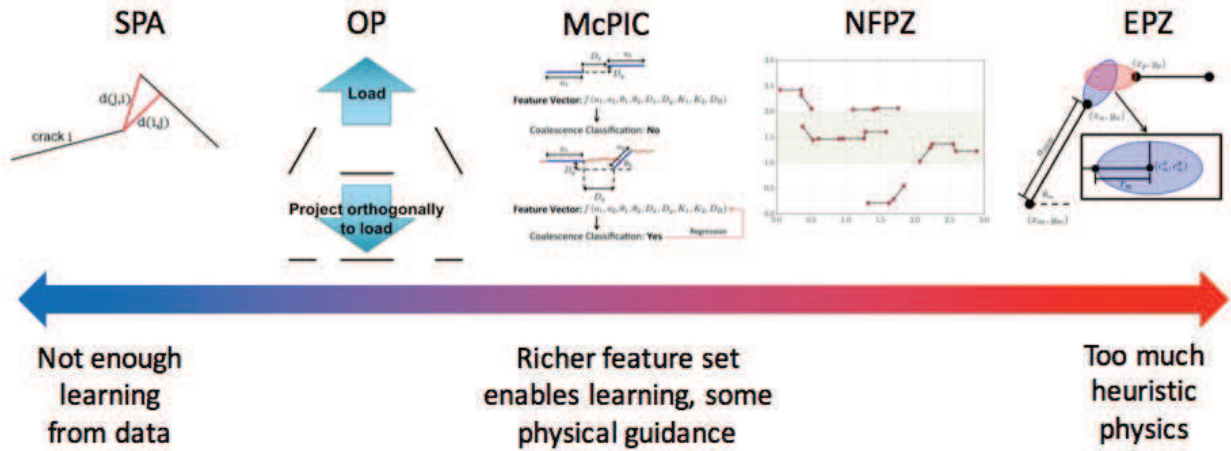}
\centering
\caption{The five models presented in this study represented on a spectrum that ranges from purely data-driven to incorporating heuristic physics information. } 
\label{fig:eight}
\end{figure*}

Comparison between these methods would be incomplete if it did not provide us with insights into the relative importance of using a combination of physics intuition and training data. 
Figure~\ref{fig:eight} shows where the different methods fall on this spectrum of training features that go from purely data-driven to physics informed constraints. 
Methods at either end of this spectrum performed poorly in terms of the metrics considered here. 
The methods on the left had an insufficient ability to learn from the training data with the SPA approach not utilizing the training data at all and the OP approach using only one feature (crack length) to predict crack growth.  
In contrast, the methods on the right included more physics, but the physics was simplified which limited their ability make accurate predictions.  
The rigid manner in which physics was incorporated in these models also made it difficult for the learning to overcome the simplifications. 
The McPIC algorithm struck a balance between these two approaches by including a rich feature set including physical information (such as the stress intensity factor) while utilizing ML algorithms to learn the behavior of the system based on these features.  
This research exercise further confirms our notion that there exists a ``Goldilocks'' region, where the judicious use of physics constraints in combination with training data is more amenable to learning the underlying complex physics of the fracture propagation problem.

%% file: Sections/S6_Summary.tex

\section{SUMMARY}
\label{sec:summary}
In summary, we have presented five different approaches (SPA, OP, McPIC, NFPZ, and EPZ models) that could be used to develop reduced-order models for informing larger length-scale simulations.  
The problem of interest focused on in this work is crack network evolution and damage accumulation in brittle geomaterials.  
In particular, we narrow the problem to address Mode I failure in a 2D concrete sample under pure tensile loading conditions.  
HOSS, the high-fidelity model chosen for this work, produced a training dataset of 150 simulations where the orientations and locations of the 20 preexisting cracks in the sample are randomly chosen for each simulation. 
The OP, McPIC, NFPZ, and EPZ methods all rely on ML algorithms to help approximate the system by attempting to learn the dominant trends and effects that can determine an overall material response.  
These methods all used the same 150 simulation training set to inform the ML algorithms.  
The SPA algorithm does not utilize ML algorithms, but provides an important baseline case for comparison.

Two metrics were chosen for comparison: the failure pathway and the failure time.  
An additional 35 HOSS simulations were used for validation of the methods.  
Of the methods, the McPIC approach produced the most promising results for both metrics used in the comparison.  
This method is specifically focused on accounting for pairwise crack interactions.  
Future work includes incorporating higher order crack interactions into this approach to better account for crack propagation and improve results.  
The method could be used to produce probability density functions of how the overall damage in the cell is changing with time.  
Such information could then be used to directly inform macro-scale continuum constitutive models through degrading elastic moduli, producing an overall material response that includes crack interaction effects in a model that cannot explicitly evolve discrete cracks. 
The key finding in this research study is that a combination of physics-informed and data-driven features are necessary for emulating the physics of fracture propagation, interaction and coalescence.